\documentclass[amsmath,amssymb,aps,twocolumn,superscriptaddress,longbibliography]{revtex4-1}

\usepackage{braket}
\usepackage{mathtools}

\usepackage[dvipsnames]{xcolor}

\usepackage[colorlinks=false]{hyperref}

\usepackage{tikz} 

\pgfdeclarelayer{bg}    
\pgfdeclarelayer{bbg}    
\pgfsetlayers{bbg,bg,main}  

\usepackage{graphicx}
\usepackage{dcolumn}
\usepackage{bm}

\begin{document}

\preprint{APS/123-QED}

\title{Towards Quantum Machine Learning with Tensor Networks}

\author{William Huggins}
\affiliation{University of California Berkeley, Berkeley, CA 94720 USA}

\author{Piyush Patil}
\affiliation{University of California Berkeley, Berkeley, CA 94720 USA}

\author{Bradley Mitchell}
\affiliation{University of California Berkeley, Berkeley, CA 94720 USA}

\author{K. Birgitta Whaley}
\affiliation{University of California Berkeley, Berkeley, CA 94720 USA}

\author{E.\ Miles Stoudenmire}
\affiliation{Center for Computational Quantum Physics, Flatiron Institute, 162 5th Avenue, New York, NY  10010, USA}

\date{\today}

\begin{abstract}
Machine learning is a promising application of quantum computing, but challenges remain
as near-term devices will have a limited number of physical qubits and high error rates.
Motivated by the usefulness of tensor networks for machine learning in the classical context,
we propose quantum computing approaches to both discriminative and generative learning, 
with circuits based on tree and matrix product state tensor networks that could have benefits
for near-term devices. The result is
a unified framework where classical and quantum computing can
benefit from the same theoretical and algorithmic developments, and the same model can
be trained classically then transferred to the quantum setting for additional optimization.
Tensor network circuits can also provide qubit-efficient schemes where, depending on the architecture,
the number of physical qubits required scales only logarithmically with, or independently of the 
input or output data sizes. We demonstrate our proposals with numerical experiments,
training a discriminative model to perform handwriting recognition using a optimization 
procedure that could be carried out on quantum hardware, and testing the noise resilience of the 
trained model.
\end{abstract}

\maketitle

\section{\label{sec:introduction} Introduction}

For decades, quantum computing has promised to revolutionize certain computational tasks. 
It now appears that we stand on the eve of the first experimental demonstration of a 
quantum advantage \cite{Boixo:2016}. 
With noisy, intermediate scale quantum computers around the corner, it is natural to investigate the 
most promising applications of quantum computers and to determine how best to harness the limited, yet
powerful resources they offer.

Machine learning is a very appealing application for quantum computers because the theories of learning and of quantum mechanics both involve statistics at a fundamental level, and machine learning techniques are  inherently resilient to noise, which may allow realization by near-term quantum computers operating without error correction. But major obstacles include the limited number of qubits in near-term devices and the challenges of working with real data. Real data sets may contain millions of samples, and individual 
samples are typically vectors with hundreds or thousands of components. 
Therefore one would like to find quantum algorithms that can perform meaningful tasks 
for large sets of high-dimensional samples even with a small number of noisy qubits.

The quantum algorithms we propose in this work implement machine learning tasks---both
discriminative and generative---using circuits equivalent to tensor networks 
\cite{Ostlund:1995,Orus:2014a,Verstraete:2008}, specifically tree tensor networks 
\cite{Fannes:1992,Lepetit:2000,Tagliacozzo:2009,Hackbusch:2009}
and matrix product states \cite{Ostlund:1995,Vidal:2003,Schollwoeck:2011}.
Tensor networks have recently been proposed as a promising architecture for
machine learning with classical computers \cite{Cohen:2015,Novikov:2016,Stoudenmire:2016s}, 
and provide good results for both discriminative \cite{Novikov:2016,Stoudenmire:2016s,Levine:2017,Liu:2017,Khrulkov:2017,Stoudenmire:2018,Glasser:2018} and generative learning tasks \cite{Han:2017,Guo:2018}.

The circuits we will study contain many parameters which are not 
determined at the outset, in contrast to quantum algorithms such as 
Grover search or Shor factorization \cite{Grover:1996,Shor:1997}.  
Only the circuit geometry is fixed, while the parameters determining the unitary 
operations must be optimized for the specific machine learning task.
Our approach is therefore conceptually related to the quantum variational 
eigensolver \cite{Peruzzo:2014,McClean:2016} and to the quantum approximate optimization algorithms
\cite{Farhi:2014}, where quantum circuit parameters are discovered with the help of an auxiliary
classical algorithm.

\begin{figure}[t]
\includegraphics[width=\columnwidth]{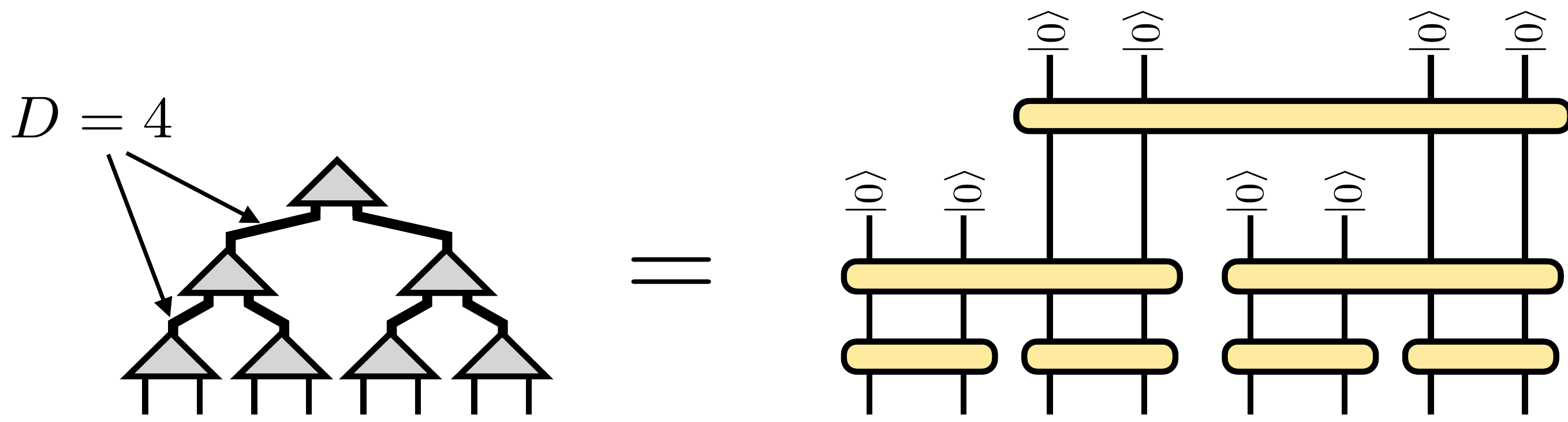}
\caption{The quantum state of $N$ qubits corresponding to a tree tensor network (left) can be realized 
as a quantum circuit acting on $N$ qubits (right). The circuit is read from top to bottom, with the yellow bars representing unitary gates. The bond dimension $D$ connecting two
nodes of the tensor network is determined by number of qubits $V$
connecting two sequential unitaries in the circuit, with $D=2^V$.}
\label{fig:tree_circuit}
\end{figure}

The application of such hybrid quantum-classical algorithms to machine learning was 
recently investigated by several groups for labeling \cite{Farhi:2018,Schuld:2018} or generating 
data \cite{Gao:2017,Benedetti:2018,Mitarai:2018}. The proposals of 
Refs.~\onlinecite{Farhi:2018,Benedetti:2018,Mitarai:2018,Schuld:2018} are related to approaches
we propose below, but consider very general classes of quantum circuits. 
This motivates the question: is there a subset of
quantum circuits  which are especially natural or advantageous for machine learning tasks? 
Tensor network circuits might provide a compelling answer, for three main reasons:

\begin{enumerate}

\item Tensor network models could be implemented on \textbf{small, near-term quantum devices}
for input and output dimensions far exceeding the number of physical qubits. 
If the hardware permits the measurement of one of the qubits separately from the others,
then the number of physical qubits needed can be made to scale either logarithmically with  
the size of the processed data, or independently of the data size depending on the particular
tensor network architecture. Models based on tensor networks may also have an inherent resilience to
noise. We explore both of these aspects in Section~\ref{sec:near}.

\item There is a \textbf{gradual crossover} from classically simulable tensor network circuits to 
circuits that require a quantum computer to evaluate. 
With classical resources, tensor network models
already give very good results for supervised \cite{Novikov:2016,Stoudenmire:2016s,Liu:2017,
Stoudenmire:2018}
and unsupervised \cite{Han:2017,Stoudenmire:2018} learning tasks. 
The same models---with the same dataset size
and data dimension---can be used to
initialize more expressive models requiring quantum hardware, 
making the optimization of the quantum-based model faster and 
more likely to succeed. Algorithmic improvements in the classical setting
can be readily transferred to the quantum setting as well.

\item There is a rich \textbf{theoretical understanding} of the properties of tensor networks 
\cite{Orus:2014a,Verstraete:2008,Schollwoeck:2011,Evenbly:2011g,Hastings:2007,Ostlund:1995}, 
and their relative mathematical simplicity (involving only linear operations) 
will likely facilitate further conceptual  developments in the machine learning
context, such as interpretability and generalization.
Properties of tensor networks, such as locality of correlations, may provide
a favorable inductive bias for processing natural data \cite{Levine:2017}.
One can prove rigorous bounds on the noise-resilience of quantum circuits based
on tensor networks \cite{Kim:2017}.

\end{enumerate}

All of the experimental operations necessary to implement tensor network circuits are available for
near-term quantum hardware. The capabilities required are preparation of product states;
one- and two-qubit unitary operations; and measurement 
in the computational basis.

In what follows, we first describe our proposed frameworks for discriminative and generative learning tasks
in Section~\ref{sec:learning}.
Then we present results of a numerical experiment which demonstrates the feasibility
of the approach using operations that could be carried out with an actual quantum device in 
Section~\ref{sec:expts}.
We conclude by discussing how the learning approaches could be implemented with a small
 number of physical qubits and by addressing their resilience to noise in Section~\ref{sec:near}.

\section{Learning with Tensor Network Quantum Circuits \label{sec:learning}}

The family of tensor networks we will consider---tree tensor networks and 
matrix product states---can always be realized precisely by a quantum 
circuit; see Fig.~\ref{fig:tree_circuit}.
Typically, the quantum circuits corresponding to tensor networks are carefully devised to
make them efficient to prepare and manipulate with classical computers \cite{Vidal:2008}.
With increasing bond dimension, tree and matrix product state tensor gradually
capture a wider range of states, which translates into 
more expressive and powerful models within the context of machine learning.
 
For very large bond dimensions, tree and matrix product tensor networks can eventually 
encompass the entire state space. But when the bond dimensions become too high, 
the cost of the classical approach becomes prohibitive. By implementing tensor network 
circuits on quantum hardware instead, one could go far beyond the space of classically
tractable models.

In this section, we first describe our tensor-network based proposal for performing discriminative tasks
with quantum hardware. The goal of a discriminative model is to produce a specific output given
a certain class of input; for example, assigning labels to images. Then we describe
our proposal for generative tasks, where the goal is to generate samples 
from a probability distribution inferred from a data set. For more background on 
various types of machine learning tasks, see the recent review Ref.~\onlinecite{Mehta:2018}.

For clarity of presentation, we shall make use of multi-qubit unitary operations in this work.  However we recognize that in practice such unitaries must be implemented using a more limited set of few-qubit operations, such as the universal gate sets of one- and two-qubit operators.  
Whether it is more productive to classically optimize over more general unitaries then 
``compile'' these into few-qubit operations as a separate step, or to parameterize the 
models in terms of fewer operations from the outset remains an interesting and important practical 
question for further work.

\subsection{Discriminative Algorithm \label{sec:discriminative}}

To explain the discriminative tensor network framework that we propose here, 
 assume that the input to the algorithm takes the form of a vector of $N$ real numbers
\mbox{$\mathbf{x} = (x_1,x_2,\ldots,x_N)$}, with each component normalized such
that $x_i \in [0,1]$.
For example, such an input could
correspond to a grayscale image with $N$ pixels, with individual entries encoding normalized grayscale values. We map this vector \(\mathbf{x} \in \mathbb{R}^N\) to a product state on N qubits according to the feature map proposed in Ref. \onlinecite{Stoudenmire:2016s}:
\begin{align}
	\mathbf{x}\ \rightarrow & \nonumber \\
	\ \ket{\Phi(\mathbf{x})} & \!\!=\!\!
    \begin{bmatrix} \cos\big(\frac{\pi}{2} x_1\big) \\ \sin\big(\frac{\pi}{2} x_1\big) 
    \end{bmatrix}\!\!\otimes\!\!
    \begin{bmatrix} \cos\big(\frac{\pi}{2} x_2\big) \\ \sin\big(\frac{\pi}{2} x_2\big) 
    \end{bmatrix}\!\!\otimes
    \cdots \otimes\!\!
    \begin{bmatrix} \cos\big(\frac{\pi}{2} x_N\big) \\ \sin\big(\frac{\pi}{2} x_N\big) \end{bmatrix}.
    \label{eqn:feature_map}
\end{align}
Such a state can be prepared by starting from the computational basis 
state $\ket{0}^{\otimes N}$, then applying 
a single qubit unitary to 
each qubit \mbox{$n=1,2,\ldots,N$}.

\begin{figure}[t]
\includegraphics[width=\columnwidth]{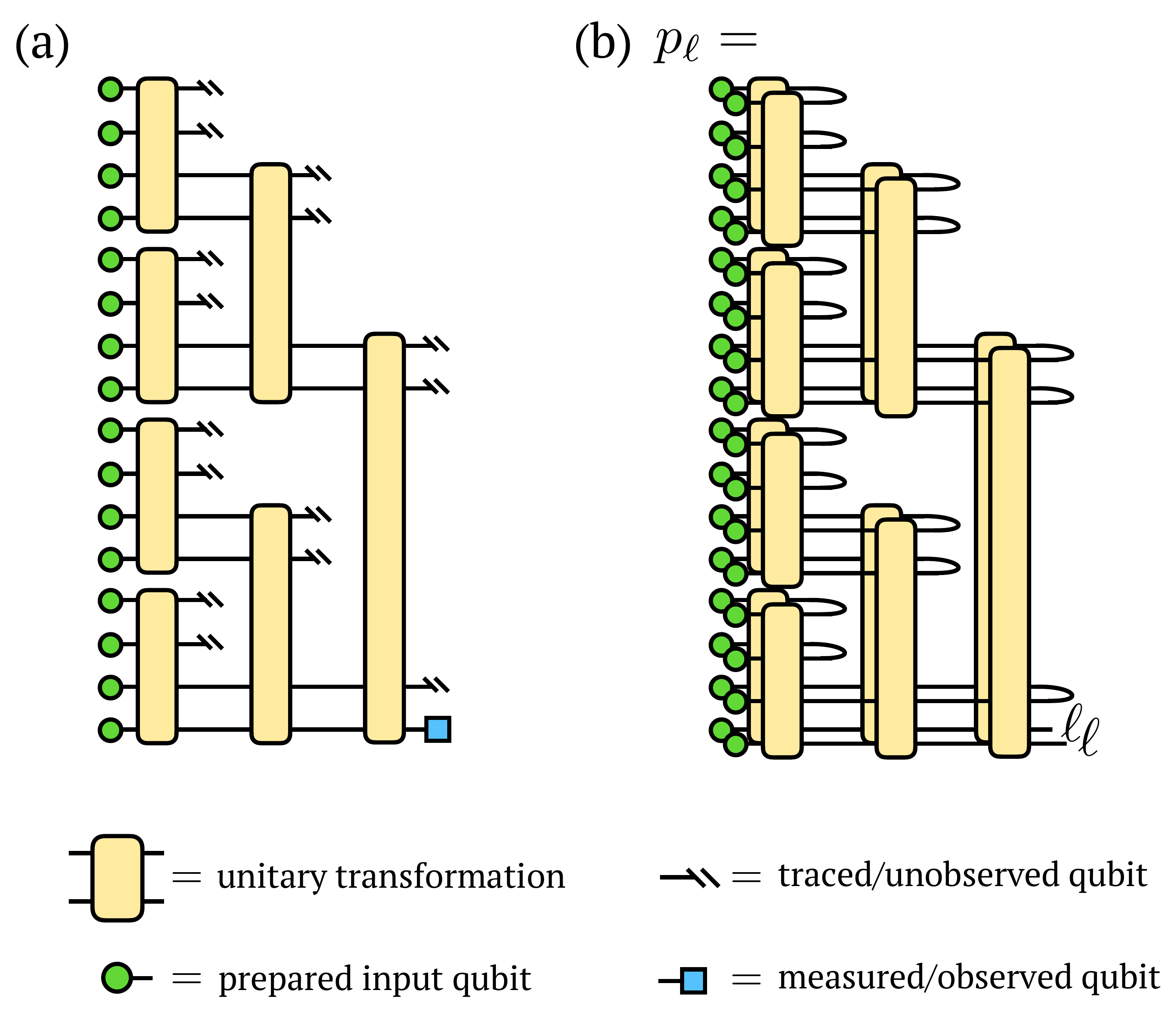}
\caption{Discriminative tree tensor network model architecture, showing
an example in which $V=2$ qubits connect different subtrees.
Figure (a) shows the model implementation as a quantum circuit. Circles indicate inputs prepared in
a product state as in Eq.~\ref{eqn:feature_map}; hash marks indicate qubits that remain
unobserved past a certain point in the circuit. A particular pre-determined qubit is
sampled (square symbol) and its distribution serves as the output of the model.
Figure (b) shows the tensor network diagram for the reduced density matrix of the output qubit.
}
\label{fig:disc_model}
\end{figure}
\begin{figure}[b]
\input{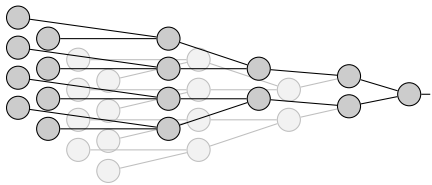}
\caption{The connectivity of nodes of our tree network model, as it would be applied to a 4x4 image. Each step coarse-grains in either the horizontal or the vertical directions, and these steps alternate.}
\label{fig:tree_schematic}
\end{figure}

The model we then propose can be seen as an iterative coarse-graining procedure that parameterizes a 
CPTP (completely positive trace preserving) map from an N-qubit input space to a small 
number of output qubits encoding the different possible class labels. 
The circuit takes the form of a tree, with $V$ qubit lines connecting each subtree to
the rest of the circuit. We call such qubit lines ``virtual qubits'' to connect
with the terminology of tensor networks, where tensor indices internal to the network are called
virtual indices. A larger $V$ can capture a larger set of functions, just as a tensor network
with a sufficiently large bond dimension can parameterize any N-index tensor. 

At each step, we take $V$ of the qubits resulting from one of the unitary operations of 
the previous step, or subtree, and $V$ from another subtree and act on them with another parameterized unitary transformation (possibly together with some ancilla qubits---not shown). 
Then $V$ of the qubits are discarded, while the other $V$ proceed to the next node of the tree,
that is, the next step of the circuit. In our classical simulations we trace over all discarded qubits, while on a quantum computer, we would be free to ignore or reset such qubits.

\begin{figure}[b]
\includegraphics[width=0.6\columnwidth]{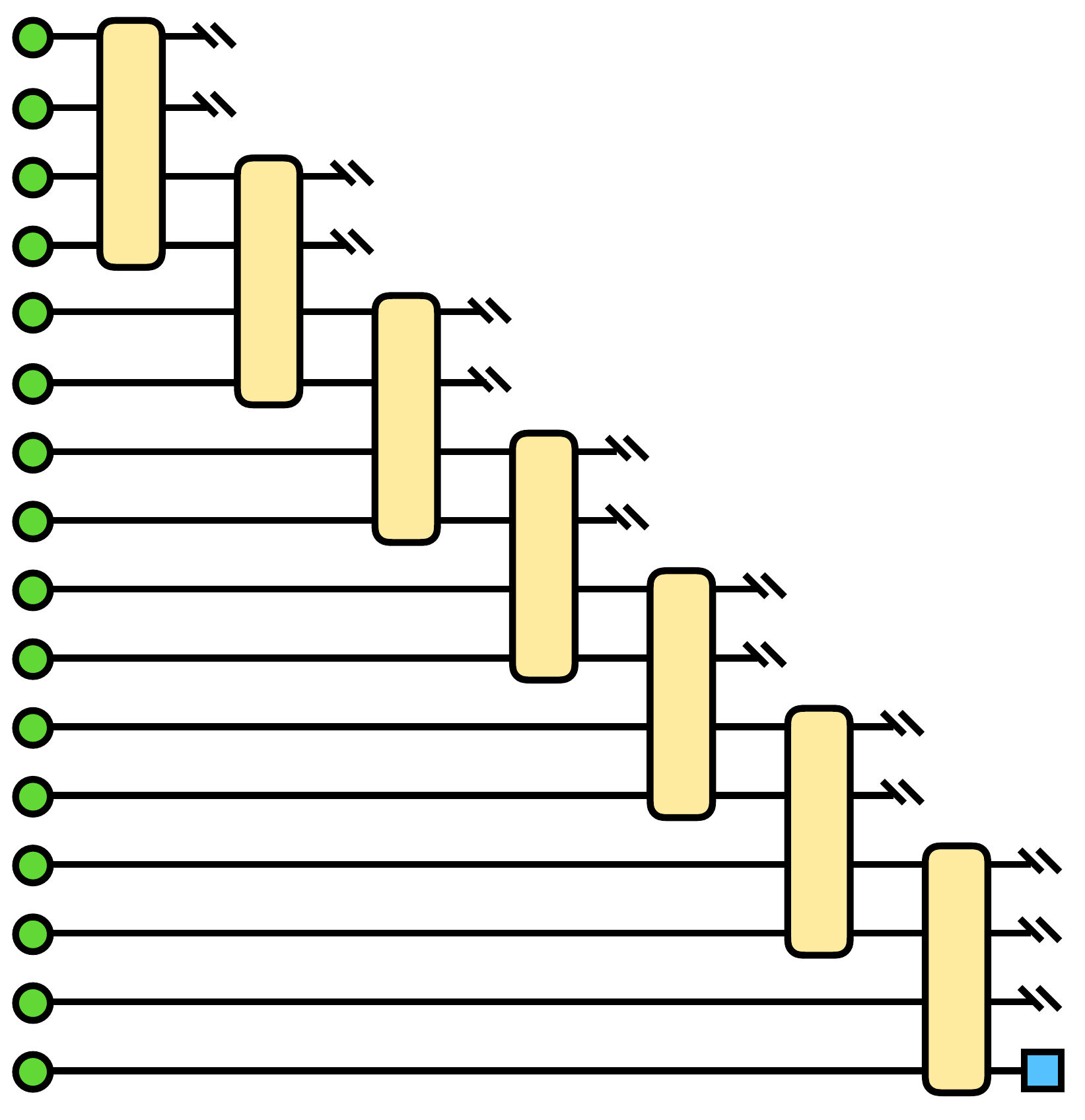}
\caption{Discriminative tensor network model 
for the case of a matrix product state (MPS) architecture
with $V=2$ qubits connecting each subtree. The symbols
have the same meaning as in Fig.~\ref{fig:disc_model}.
An MPS can be viewed as a maximally unbalanced tree.
}
\label{fig:disc_model_mps}
\end{figure}

Once all unitary operations defining the circuit have been
carried out, one or more qubits serve as the output qubits. 
(Which qubits are outputs is designated ahead of time.) 
The most probable state
of the output qubits determines the prediction of the model, that is, the label the model assigns
to the input. To determine the most probable state of the output qubits, 
one performs repeated evaluations of the circuit for the same input in order to 
estimate their probability distribution in the computational basis. 

We show the quantum circuit of our proposed procedure in Fig.~\ref{fig:disc_model}.
In the case of image classification, it is natural to always group input qubits based on pixels coming from nearby regions of the image, with a tree structure illustrated schematically in Fig. \ref{fig:tree_schematic}. 

A closely related family of models can be devised based on matrix product states. 
An example is illustrated in Fig.~\ref{fig:disc_model_mps} showing the case of $V=2$. 
Matrix product states (MPS) can be viewed as maximally unbalanced trees, and differ from the binary tree 
models described above in that after each unitary operation on $2V$ inputs 
only one set of $V$ qubits are passed to the next node of the network. Such models
are likely a better fit for data that has a one-dimensional pattern of correlations, such as 
time-series, language, or audio data.

\subsection{Generative Algorithm}
\begin{figure}[b]
\includegraphics[width=0.5\columnwidth]{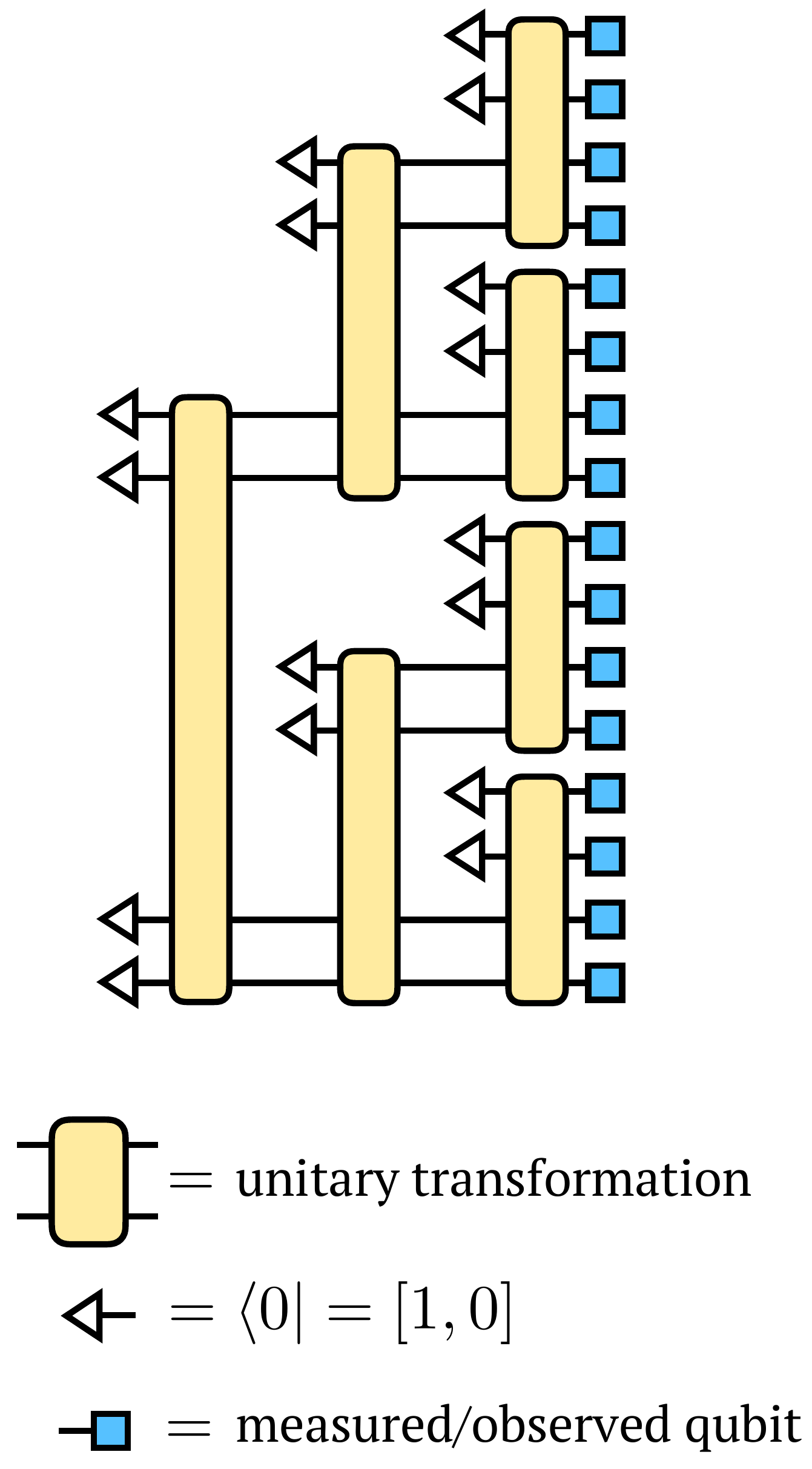}
\caption{Generative tree tensor network model architecture, showing a case
with $V=2$ qubits connecting each subtree.
To sample from the model, qubits are prepared in a reference computational
basis state $\bra{0}$ (left-hand side of circuit). Then $2V$ qubits are entangled via unitary operations
at each layer of the tree as shown. The qubits are measured at the points in the
circuit labeled by square symbols (right-hand side of circuit), 
and the results of these measurements provides
the output of the model. While all qubits could be entangled before being
measured, we discuss in Section~\ref{sec:near} the possibility performing
opportunistic measurements to reduce the physical qubit overhead.}
\label{fig:gen_model}
\end{figure}
\begin{figure}[b]
\includegraphics[width=0.6\columnwidth]{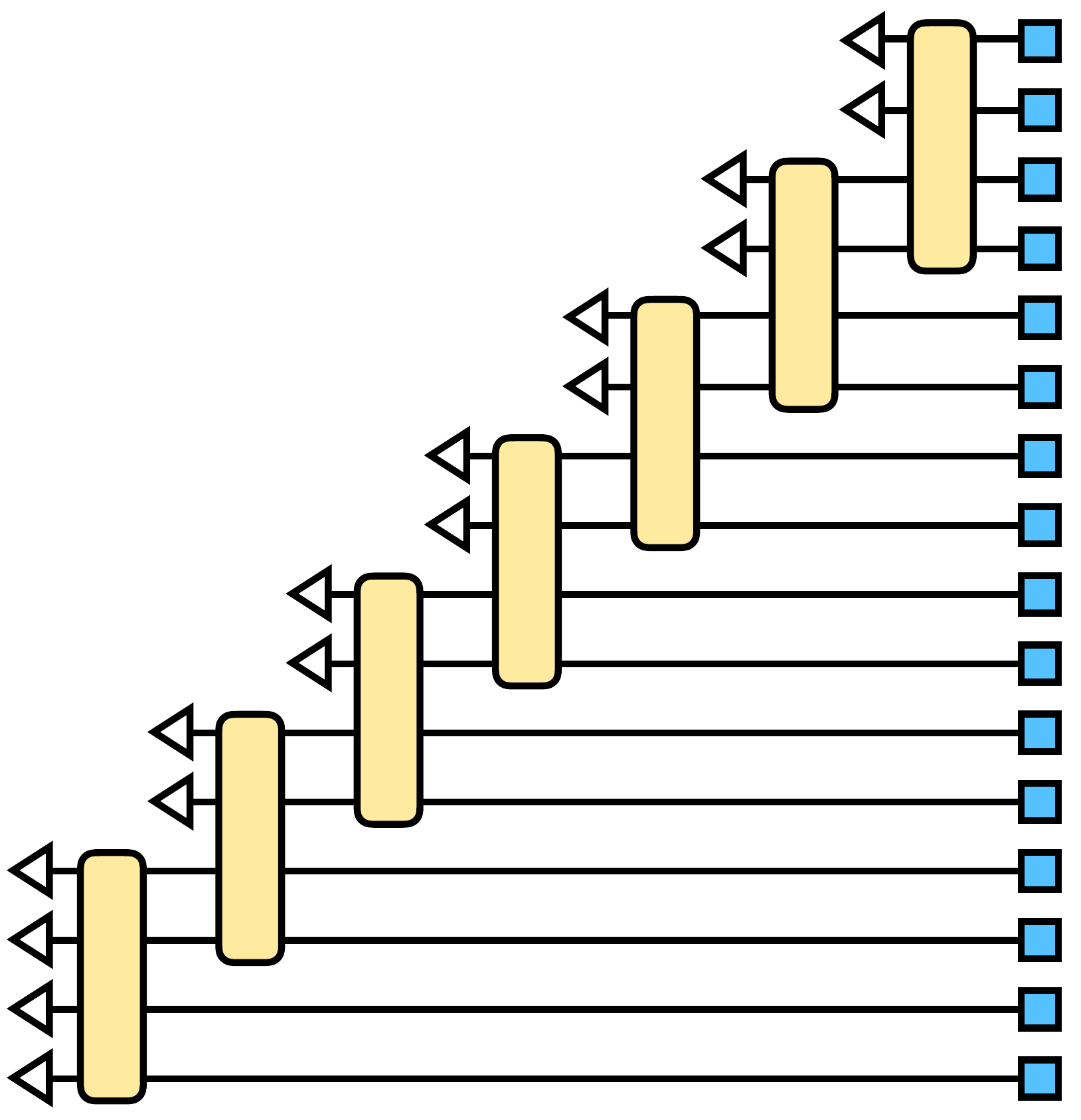}
\caption{Generative tensor network model 
for the case of a matrix product state (MPS) architecture
with $V=2$ qubits connecting each unitary. The symbols
have the same meaning as in Fig.~\ref{fig:gen_model}.}
\label{fig:gen_model_mps}
\end{figure}

The generative algorithm we propose is nearly the reverse of the discriminative algorithm,
in terms of its circuit architecture. The algorithm produces random samples by first preparing
a quantum state then measuring it in the computational basis, putting it within the family 
of algorithms  recently dubbed ``Born machines'' \cite{Han:2017,Gao:2017,Benedetti:2018}. 
But rather than preparing a completely general state, we shall consider 
specific patterns of state preparation corresponding to tree and matrix product state tensor networks.
This provides the advantages discussed in the introduction, such as
connections to classical tensor network models and the ability to reduce the number of 
physical qubits required, which will be discussed further in Section~\ref{sec:near}.

The generative algorithm based on a tree tensor network (shown in Fig.~\ref{fig:gen_model})
begins by preparing $2V$ qubits in a reference computational
basis state $\bra{0}^{\otimes 2V}$, then entangling these qubits by unitary operations.
Another set of $2V$ qubits are prepared in the state $\bra{0}^{\otimes 2V}$. Half
of these  are grouped with the first $V$ entangled qubits, and half with the second $V$ entangled
qubits. Two more unitary operations are applied to each new
grouping of $2V$ qubits; the outputs are now split into four groups; and the process
repeats for each group. The process ends when the total number of qubits processed 
reaches the size of the output one wants to generate.

Once all unitaries acting on a certain qubit have been applied, this qubit can
be measured. The measured output of all of the qubits in the computational 
basis represents one sample from the generative model. 

We illustrate our proposed
generative approach for the case of $V=2$ and binary outputs in Fig.~\ref{fig:gen_model}.
As in the discriminative case, one can also devise an MPS based generative algorithm 
more suitable for one-dimensional data. The circuit for such an algorithm is shown
in Fig.~\ref{fig:gen_model_mps}.

\section{Numerical Experiments \label{sec:expts}}

\begin{figure}[t]
\includegraphics[width=0.5\columnwidth]{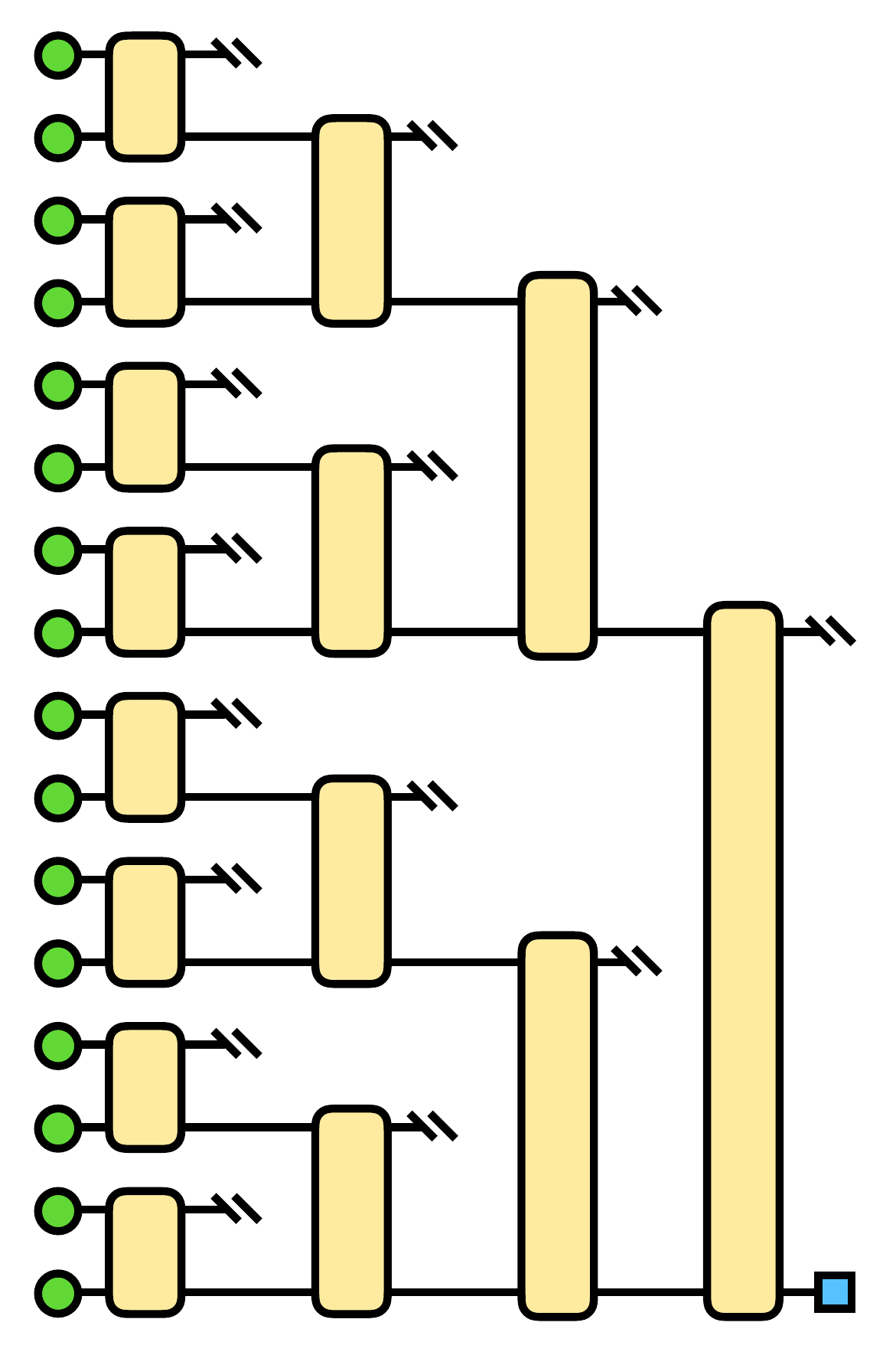}
\caption{Model architecture used in the experiments of Section~\ref{sec:expts},
which is a special case of the model of Fig.~\ref{fig:disc_model} with one
virtual qubit connecting each subtree. For illustration purposes we show a model
with 16 inputs and 4 layers above, whereas the actual model used in the experiments
had 64 inputs and 6 layers.}
\label{fig:d2_tree}
\end{figure}

To show the feasibility of implementing our proposal on a near-term quantum device,
we trained a discriminative model based on a tree tensor
network for a supervised learning task, namely labeling image data. 
The specific network architecture we used is shown as a quantum circuit 
in Fig.~\ref{fig:d2_tree}. When viewed as a tensor network, this model
has a bond dimension of $D=2$. This stems from the fact that after each
unitary operation entangles two qubits, only one of the qubits is acted on at the next scale
(next step of the circuit).

\subsection{Loss Function}

Our eventual goal is to select the parameters of our circuit such that we can confidently assign the correct label to a new piece of data by running our circuit a small number of times.  To this end, we choose the
 loss function which we want to minimize starting with the following definitions.
Let \(\mathbf{\Lambda}\) be the model parameters; \(\mathbf{d}\) be an element of the training data set;
and let $p_\ell(\mathbf{\Lambda},\mathbf{x})$ be the probability of the model to output a label $\ell$ for a given input $\mathbf{x}$.
Because we consider the setting of supervised learning, the correct labels are known for the 
training set inputs, and define $\ell_\mathbf{x}$ to be the correct label for the input $\mathbf{x}$.
Now define 
\begin{align}
p_\text{largest false}(\mathbf{\Lambda}, \mathbf{x}) = \max_{\ell \neq \ell_\mathbf{x}}\Big[ p_\ell(\mathbf{\Lambda},\mathbf{x}) \Big]
\end{align}
as the probability of the incorrect output state which has the highest probability of being
observed.
Then,  define the loss function for a single input $\mathbf{x}$ to be
\begin{align}
	L(\mathbf{\Lambda}, \mathbf{x}) = \max(p_\text{largest false}(\mathbf{\Lambda}, \mathbf{x}) - p_{\ell_\mathbf{x}}(\mathbf{\Lambda}, \mathbf{x}) + \lambda, 0)^\eta,
    \label{eq:loss_per_datum}
\end{align}
and the total loss function to be
\begin{align}
	L(\mathbf{\Lambda}) = \frac{1}{|\text{data}|}\sum_{\mathbf{x} \in \text{data}} L(\mathbf{\Lambda}, \mathbf{x}).
    \label{eq:loss_averaged_over_all}
\end{align}
The ``hyper-parameters'' \(\lambda\) and \(\eta\) are to be chosen to give good empirical performance
on a validation data set. 
Essentially, we assign a penalty for each element of the training set where the gap between probability of assigning the true label and the probability of assigning the most likely incorrect label is 
less than \(\lambda\). This loss function allows us to concentrate our efforts during training on making sure that we are likely to assign the correct label after taking the majority vote of several executions of the model, rather than trying to force the model to always output the correct label in each separate run.

\subsection{Optimization}

Of course, we are interested in training our circuit to generalize well to unobserved inputs, so instead of optimizing over the entire distribution of data as in Eq. \ref{eq:loss_averaged_over_all}, we optimize the loss function over a subset of the training data and compare to a held-out set of test data. Furthermore, because the size of the training set for a typical machine learning problem is so large (60,000 examples in the case of the MNIST data set), it would be impractical to calculate the loss over all of the training data at each optimization step. Instead, we follow a standard approach in machine learning and randomly select a mini-batch of training examples at each iteration. Then, we use the following stochastic estimate of our true training loss (recalling that \(\mathbf{\Lambda}\) represents the current model parameters):
\begin{align}
	\tilde{L}(\mathbf{\Lambda}) = \frac{1}{|\text{mini-batch}|}\sum_{\mathbf{x} \in \text{mini-batch}} L(\mathbf{\Lambda}, \mathbf{x})
    \label{eq:loss_averaged_over_mini-batch}
\end{align}

In order to faithfully test how our approach would perform on a near-term quantum computer, we have chosen to minimize our loss function using a variant of the simultaneous perturbation stochastic approximation (SPSA) algorithm which was recently used to find quantum circuits approximating ground states in Ref.~\onlinecite{Kim:2017} and was originally developed in Ref.~\onlinecite{spall1998overview}. 

Essentially, each step of SPSA estimates the gradient of the loss function by performing a finite difference calculation along a random direction and updates the parameters accordingly. In our experimentation, we have also found it helpful to include a momentum term $\mathbf{v}$, which 
mixes a fraction of previous update steps into the current update. We outline the algorithm we used in more detail below.
\begin{enumerate}
\item Initialize the model parameters \(\mathbf{\Lambda}\) randomly, and set \(\mathbf{v}\) to zero.
\item Choose appropriate values for the constants, \(a, b, A, s, t, \gamma, n, M\) that define the optimization procedure.
\item For each \(k \in \{0, 1, 2, ..., M\}\), set \(\alpha_k = \frac{a}{(k + 1 + A)^s}\) and \(\beta_k = \frac{b}{(k + 1)^t}\), and randomly partition the training data into mini-batches of \(n\) images. Perform the following steps using each mini-batch:
\begin{enumerate}
\item Generate random perturbation \(\mathbf{\Delta}\) in parameter space.
\item Evaluate \(g =  \frac{\tilde{L}(\mathbf{\Lambda}_{old} + \alpha_k \mathbf{\Delta}) - \tilde{L}(\mathbf{\Lambda}_{old} - \alpha_k \mathbf{\Delta})}{2 \alpha_k} \), with \(\tilde{L}(\mathbf{x})\) defined as in Eq.~\ref{eq:loss_averaged_over_mini-batch}.
\item Set \(\mathbf{v}_{new} = \gamma \mathbf{v}_{old} - g \beta_k \mathbf{\Delta}\)
\item Set \(\mathbf{\Lambda}_{new} = \mathbf{\Lambda}_{old} + \mathbf{v}_{new}\)

\end{enumerate}
\end{enumerate}

\begin{figure}[t]
\centering
\includegraphics[width=0.50\textwidth]{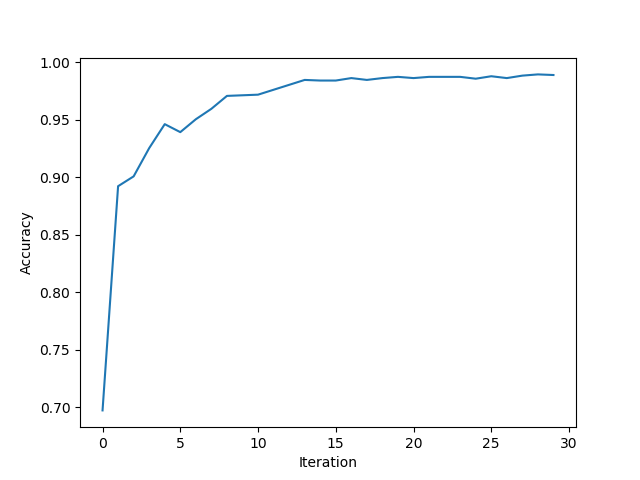}

\caption{Test accuracy as a function of the number of SPSA epochs (\(M = 30\), in the language of the previous section) for binary classification of handwritten 0's and 7's from the MNIST data set.}
\label{fig:simple_network_accuracy}
\end{figure}

\begin{figure}[b]
\centering
\includegraphics[width=0.50\textwidth]{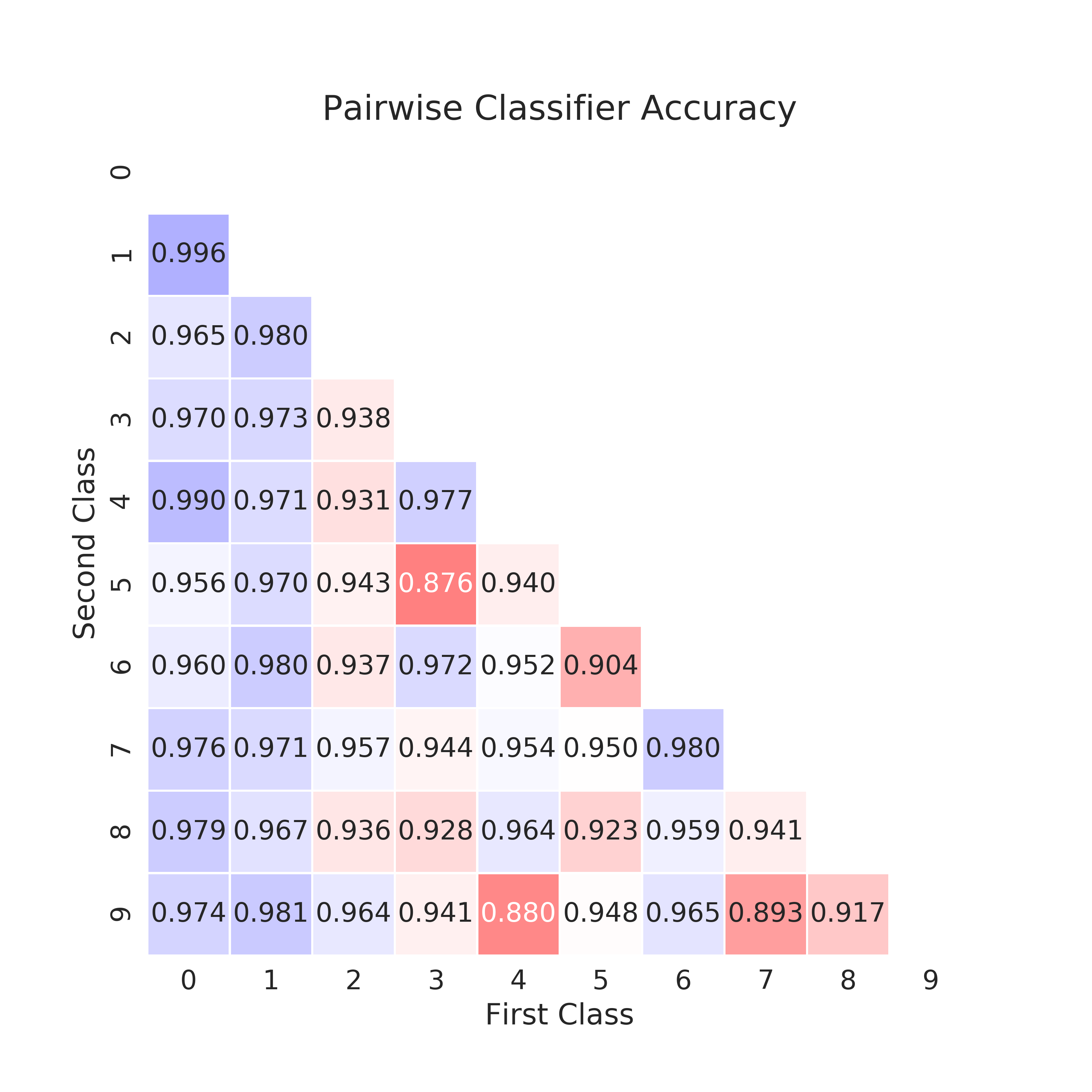}

\caption{The test accuracy for each of the pairwise classifiers trained with the
  hyper-parameters mentioned in the text. The accuracy for each classifier can
  be found by choosing the position along the x-axis corresponding to one class
  and the position on the y-axis corresponding to the other.}
\label{fig:triangle_plot_noiseless}
\end{figure}

\subsection{Results}

We trained circuits with a single output qubit at each node to recognize grayscale 
images of size $8\times 8$ belonging to one of two classes using the SPSA optimization procedure described above. The images were obtained from the MNIST data set of handwritten digits \cite{MNIST},
and we show results below for classifiers trained to distinguish between each of the 45 pairs of handwritten digits 0 through 9. 

The unitary operations $U$ applied at each node in the tree were parameterized by writing 
them as $U=\exp(i H)$ where $H$ is a Hermitian matrix (the matrices $H$ were allowed to be different
for each node). The free parameters were chosen to be
the elements forming the diagonal and upper triangle of each Hermitian matrix, resulting in exactly 1008 free parameters for the $8\times8$ image recognition task. The mini-batch size and the other hyper-parameters for the training procedure and the loss function were hand-tuned by running a small number of experiments, using the SigOpt~\cite{sigopt} software package, with the goal of obtaining the most rapid and consistent performance (averaged over the different digit pairs) on a validation data set. 

Ultimately, we found that networks trained with the choices $(\lambda = .234,\> \eta = 5.59,\> a=28.0,\> b=33.0,\> A = 74.1,\> s=4.13,\> t=.658,\> \gamma=0.882,\> n=222)$ were able to achieve an average test accuracy above 95\%. The accuracies of the individual pairwise classifiers are tabulated in Fig.~\ref{fig:triangle_plot_noiseless}, and data from a representative example of the training process for one of the easier pairs to classify is shown in~\ref{fig:simple_network_accuracy}. We observed significant differences in performance across the different pairs, partly owing, perhaps, to the difficulty of distinguishing similar digits using 64 pixel images. We also note that different choices of hyper-parameters could significantly affect which pairs were classified most accurately.

\section{Implementation on Near-Term Devices \label{sec:near}}

A key advantage of carrying out machine learning tasks with
models equivalent to tree or matrix product tensor networks is that they could be implemented
using a very small number of physical qubits.
The key requirement is that the hardware must allow 
the measurement of individual physical qubits without further disturbing the state of the 
other qubits, a capability also required for certain approaches to 
quantum error correction \cite{Corcoles:2015}. 
Below we will first discuss how the number of qubits needed to implement either a discriminative
or generative tree tensor network model can be made to scale only logarithmically in both the 
data dimension and in the bond dimension of the network.
Then we will discuss the special case of matrix product state tensor networks,
which can be implemented with a number of physical qubits that is \emph{independent}
of the input or output data dimension.

Another key advantage of using tensor network models on near-term devices could be their
robustness to noise, which will certainly be present in 
any near-term hardware. To explore the noise resilience of our models, 
we present a numerical experiment where we evaluate the model trained in Section~\ref{sec:expts} 
with random errors, and observe whether it can still produce useful results.

\subsection{Qubit-Efficient Tree Network Models}

\begin{figure}[t]
\includegraphics[width=\columnwidth]{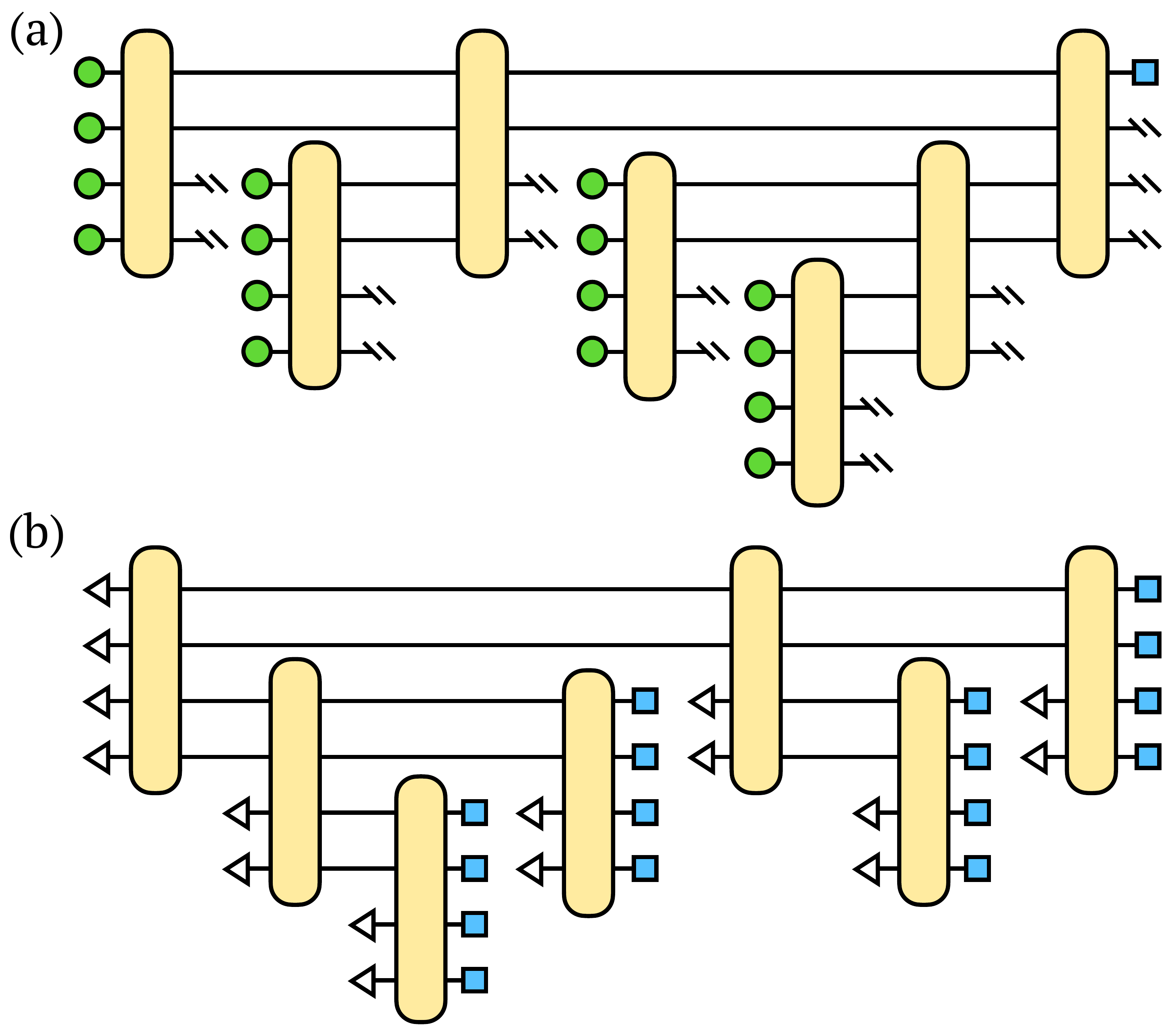}
\caption{Qubit-efficient scheme for evaluating (a) discriminative
and (b) generative tree models with $V=2$ virtual qubits and $N=16$ 
inputs or outputs. Note that the two patterns are the reverse
of each other. In (a) qubits indicated with hash marks are measured and the measurement
results discarded. These qubits are then reset and prepared with
additional input states. In (b) measured qubits are recorded and 
reset to a reference state $\bra{0}$.}
\label{fig:V2_tree}
\end{figure}

To discuss the minimum qubit resources needed to implement general tree
tensor network models, recall the notion of the virtual 
qubit number $V$ from Section~\ref{sec:learning}. 
This is the number of qubit lines connecting each
subtree to higher nodes in the tree. Viewed as a tensor network, the
bond dimension $D$, or dimension of the internal tensor indices,
is given by $D=2^V$. 

For example, the tree shown in Fig.~\ref{fig:d2_tree} has $V=1$ and
a bond dimension of $D=2$. The tree shown in Fig.~\ref{fig:V2_tree}
has $V=2$ and $D=4$. When discussing these models in general terms,
it suffices to consider only unitary operations acting on $2V$ qubits,
since at each node of the tree, two subtrees (two sets of $V$ qubits)
are entangled together.

Given only the ability to perform state preparation and unitary operations,
it would take $N$ physical qubits to evaluate a discriminative tree network
model on $N$ inputs. However, if we also allow the step of measurement and 
resetting of certain qubits, then the number of physical qubits $Q$
required to process $N$ inputs given $V$ virtual states passing between
each node can be significantly reduced to just $Q(N,V) = V \lg(2N/V)$.

To see why, consider the circuit showing the most qubit-efficient scheme for
implementing the discriminative case Fig.~\ref{fig:V2_tree}(a).
For a given $V$, the number of inputs that can be processed
by a single unitary is $2V$. Then $V$ of the qubits can be measured and reused,
but the other $V$ qubits must remain entangled. So only $V$ new qubits must
be introduced to process $2V$ more inputs. From this line of reasoning
and the observation that $Q(2V,V) = 2V$, 
one can deduce the result \mbox{$Q(N,V) = V \lg(2N/V)$}.

For generative tree network models, generating $N$ outputs
with $V$ virtual qubits requires the same number of physical qubits as 
for the discriminative case; this can be seen by observing that the
pattern of unitaries is just the reverse of the discriminative case for
the same $N$ and $V$.
Fig.~\ref{fig:V2_tree} shows the most qubit-efficient way to sample a
generative tree models for the case of $V=2$ virtual and $N=16$ output qubits,
requiring only $Q=8$ physical qubits.

Though a linear growth of the number of physical qubits as a function of virtual qubit number 
$V$ may seem more prohibitive compared to the logarithmic scaling with $N$, 
even a small increase in $V$ would lead to a significantly more expressive model.
From the point of view of tensor 
networks the expressivity of the model is usually measured by the bond 
dimension $D=2^V$. In terms of the bond dimension, the number of
qubits needed thus scales only as \mbox{$Q(N,D) \sim \lg(D)\lg(N)$}.
The largest bond dimensions used in state-of-the-art classical
tensor network calculations are around $D=2^{15}$ or about $30,000$. So for $V=16$
or more virtual qubits one would quickly exceed the power of any 
classical tensor network calculation we are aware of.

\subsection{Qubit-Efficient Matrix Product Models}

\begin{figure}[t]
\includegraphics[width=0.9\columnwidth]{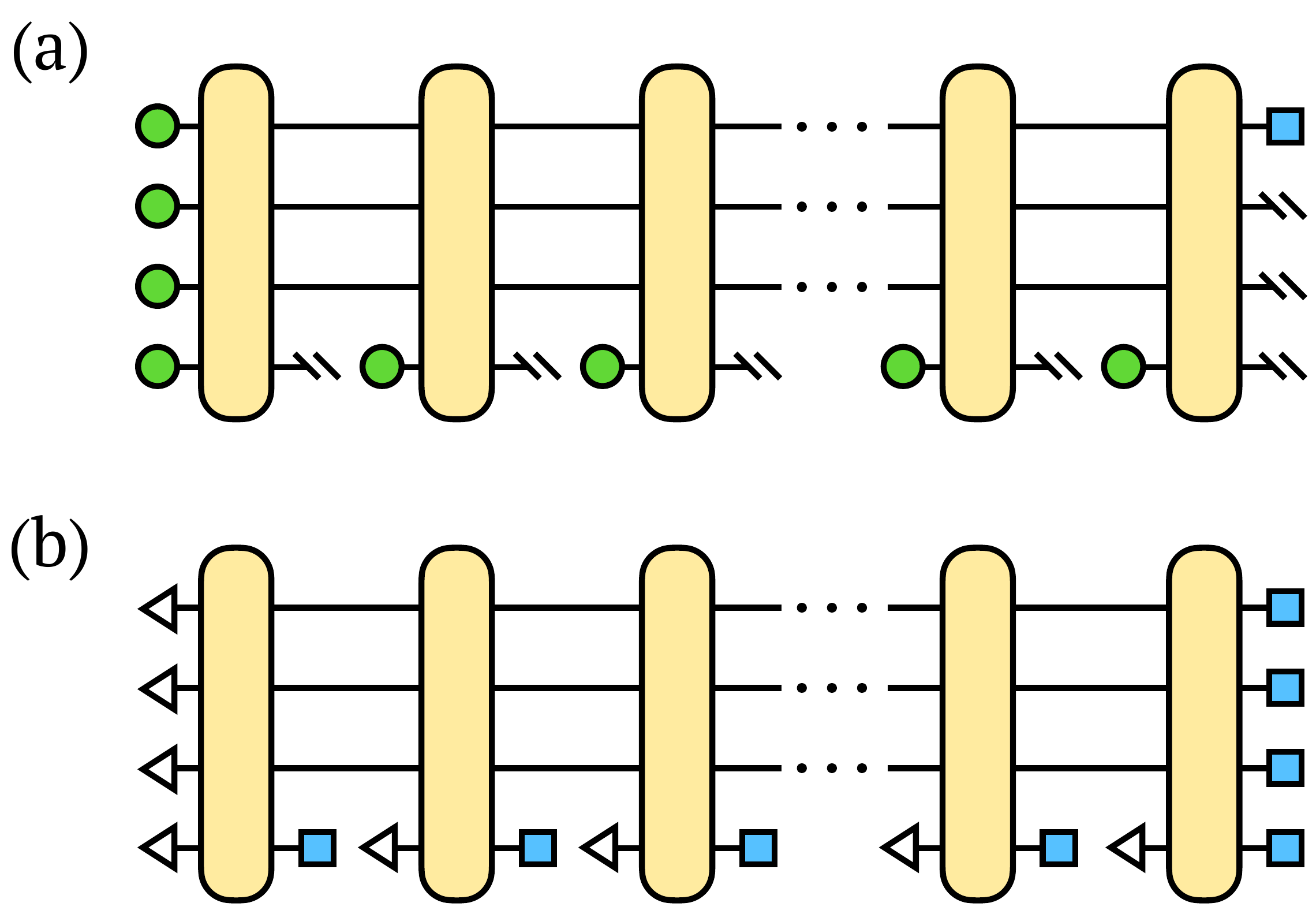}
\caption{Qubit-efficient scheme for evaluating (a) discriminative
and (b) generative matrix product state models for an arbitrary number of inputs or outputs.
The figure shows the case of $V=3$ qubits connecting each node of the network.
When evaluating the discriminative model, one of the qubits is measured after
each unitary is applied and the result discarded; the qubit is then prepared
with the next input component. To implement the generative model,
one of the qubits is measured after each unitary operation and the result recorded.
The qubit is then reset to the state $\bra{0}$.}
\label{fig:V3_mps}
\end{figure}

A matrix product state (MPS) tensor network is a special case of a tree tensor
network that is maximally unbalanced. This gives an MPS certain 
advantages without sacrificing expressivity for one-dimensional distributions,
as measured by the maximum entanglement entropy it can carry across bipartitions
of the input or output space, meaning a division of $(x_1,\ldots,x_j)$ from
$(x_{j+1},\ldots,x_N)$.

Given the ability to measure and reset a subset of physical qubits, a key
advantage of implementing a discriminative or generative tensor network
model based on an MPS is that for a model with $V$ virtual qubits, an
\emph{arbitrary} number of inputs or outputs can be processed by using only
$V+1$ physical qubits. The circuits illustrating how this can be done 
are shown in Fig.~\ref{fig:V3_mps}.

The implementation of the discriminative algorithm shown in
Fig.~\ref{fig:V3_mps}(a) begins by preparing and entangling $V$ input
qubit states. One of the qubits is measured and reset to the next input state.
Then all $V+1$ qubits are entangled and a single qubit measured and re-prepared.
Continuing in this way, one can process all of the inputs. Once all inputs are processed,
the model output is obtained by sampling one or more of the physical qubits. 

To implement the generative MPS algorithm shown in Fig.~\ref{fig:V3_mps}(b), one prepares all 
qubits to a reference state $\ket{0}^{\otimes V+1}$ and after entangling the qubits, one
measures and records a single qubit to generate the first output value. This 
qubit is reset to the state $\ket{0}$ and all the qubits are then acted on by another
$(V+1)$ qubit unitary. A single qubit is again measured to generate the second output value,
and the algorithm continues until $N$ outputs have been generated.

\begin{figure}[t]
\includegraphics[width=\columnwidth]{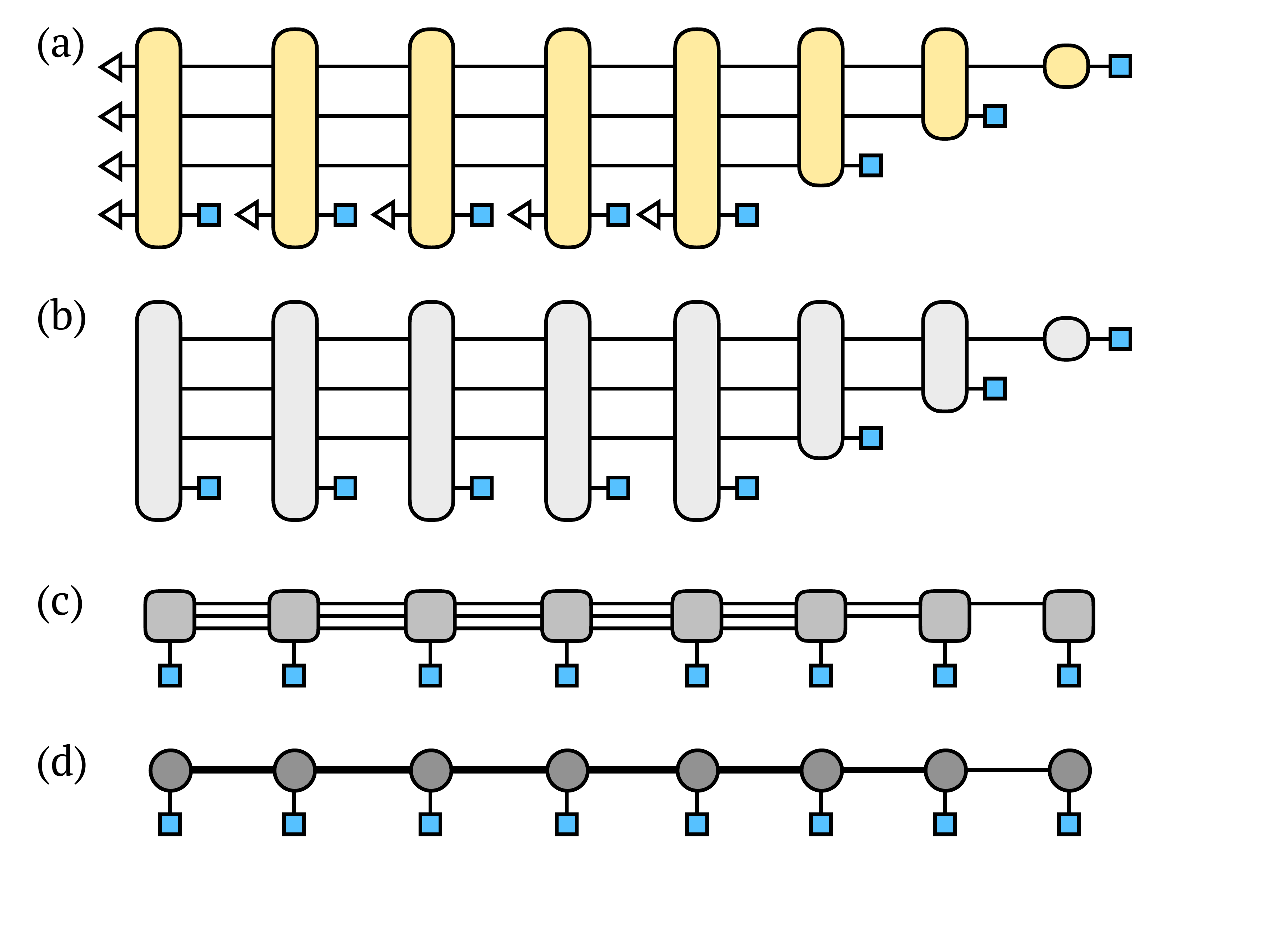}
\caption{Mapping of the generative matrix product state (MPS)
quantum circuit with $V=3$ to a bond dimension $D=2^3$ MPS
tensor network diagram. First (a) interpret the circuit diagram 
as a tensor diagram by interpreting reference states $\bra{0}$ as vectors $[1,0]$;
qubit lines as dimension 2 tensor indices; and measurements as setting
 indices to fixed values. Then (b) contract the reference states into
the unitary tensors and (c) redraw the tensors in a linear chain. Finally,
(d) merge three $D=2$ indices into a single $D=8$ dimensional 
index on each bond.}
\label{fig:gen_mps}
\end{figure}

To understand the equivalence of the generative circuit of Fig.~\ref{fig:V3_mps}(b) to 
conventional tensor diagram notation for an MPS, interpret the circuit diagram 
Fig.~\ref{fig:gen_mps}(a) as a tensor network diagram, 
treating elements such as reference states $\bra{0}$ as tensors 
or vectors $[1,0]$. One can contract or sum over the reference state indices and 
merge any $V$ qubit indices into a single index of dimension $D=2^V$. The result
is a standard MPS tensor network diagram Fig.~\ref{fig:gen_mps}(d) for the amplitude
of observing a particular set of values of the measured qubits.

\begin{figure}[b]
\centering
\includegraphics[width=0.5\textwidth]{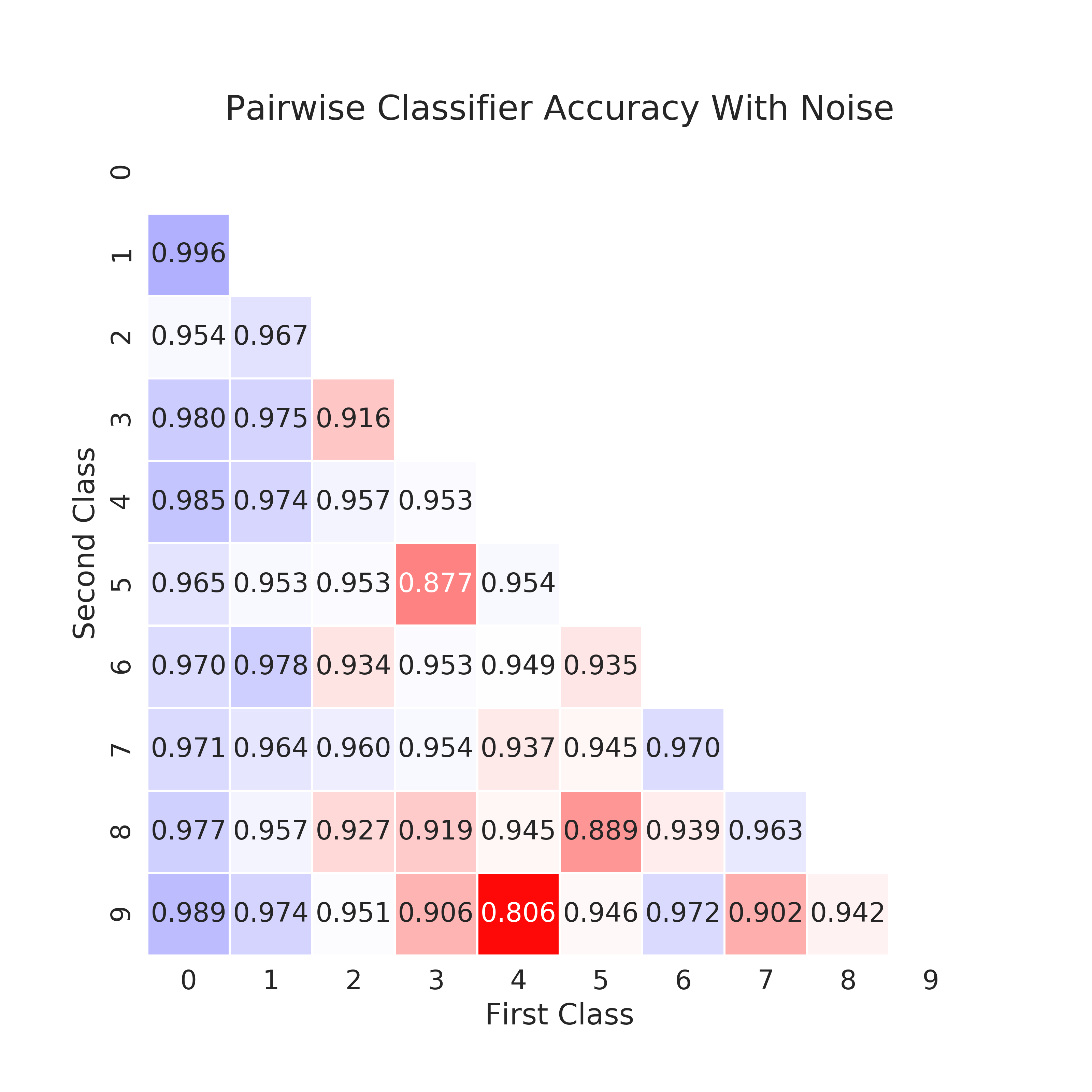}
\caption{The test accuracy for each of the pairwise classifiers under noise corresponding to a $T_1 $ of $5\,\mathrm{\mu s}$, a $T_2$ of $7\,\mathrm{\mu s}$, and a gate time of $200\,\mathrm{ns}$. In most cases, the accuracy is comparable to the results from training without noise. Note that it was necessary to choose a different set of hyper-parameters to enable successful training under noise.}
\label{fig:triangle_plot_noisy}
\end{figure}

\begin{figure}[t]
\centering
\includegraphics[width=0.5\textwidth]{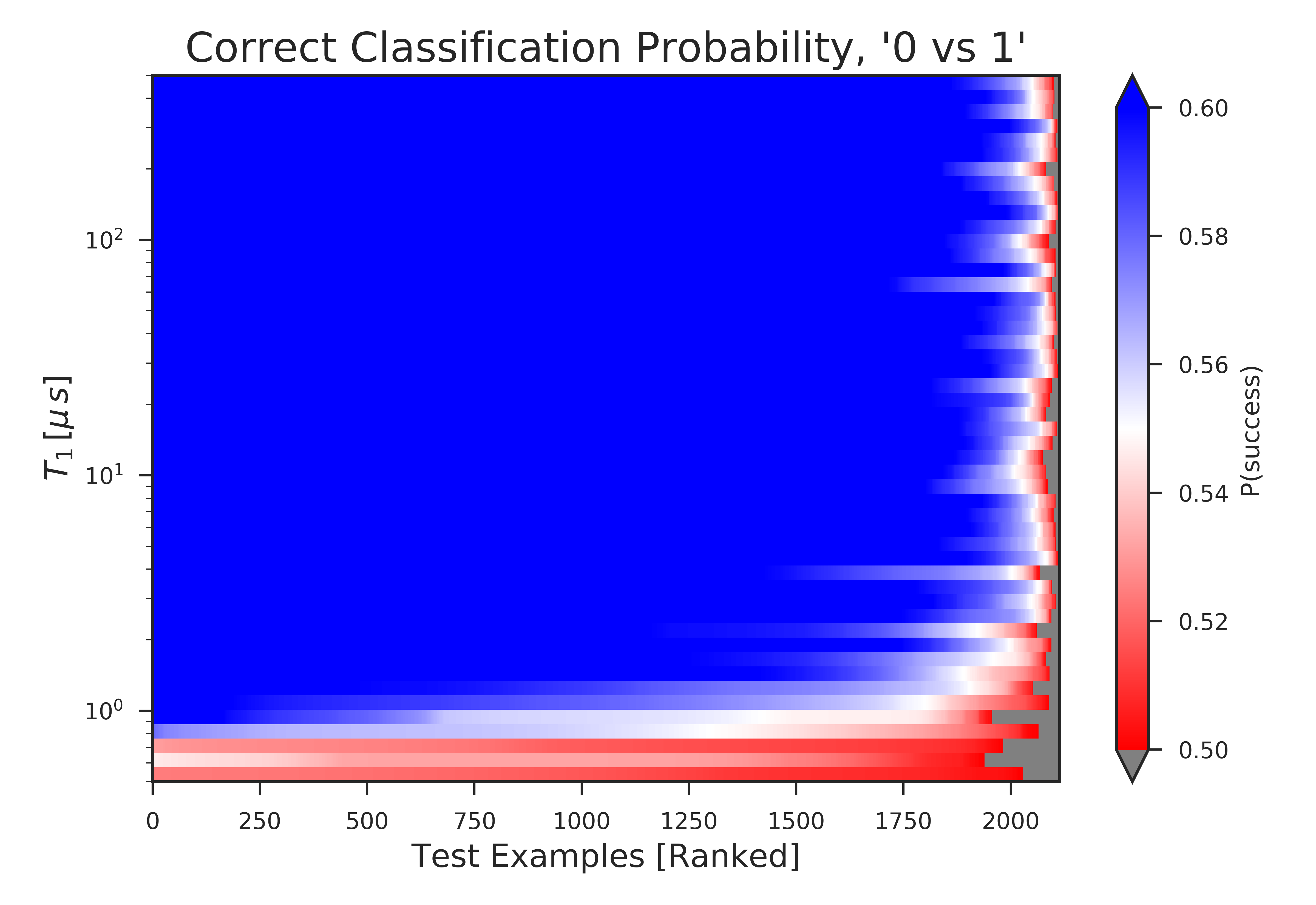}
\includegraphics[width=0.5\textwidth]{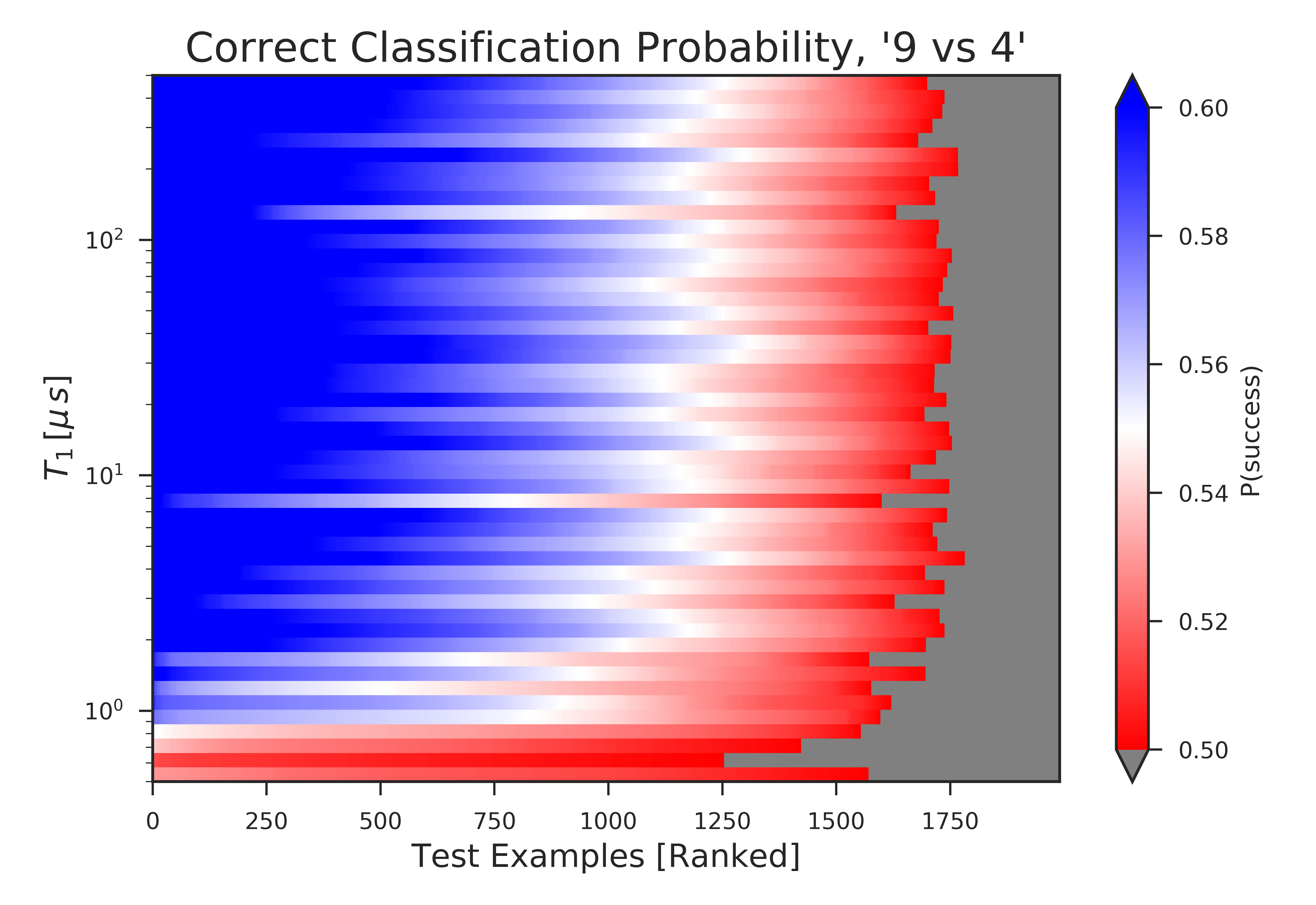}
\caption{Success probability of two different pairwise classification circuits  prediction on their test sets (sorted by decreasing probability of success along the $x$-axis) over a wide range of $T_1$ values ($y$-axis). For each \(T_1\) shown, the probability of successfully classifying each member of the test set is indicated. Note that success probabilities which are larger than \(.5\) even by a relatively small margin imply that the corresponding test example could be correctly classified with a majority voting scheme. Gate time $T_g = 200\,ns$ was held fixed while \(T_2\) was set to be \(\frac{7}{5}T_1\). Noise levels corresponding to current hardware are approximately two thirds of the way up the chart. Grey areas indicate regions where the model would misclassify the test example.}
\label{fig:noisy_evaluation}
\end{figure}

\subsection{Noise Resilience}
Any implementation of our proposed approach on near-term quantum hardware will
have to contend with a significant level of noise due to qubit and gate
imperfections. 
But one intuition about noise effects in our tree models is that an error which corrupts a qubit
only scrambles the information coming from the patch of inputs belonging to the past ``causal cone''
of that qubit.
And because the vast majority of the operations occur near the leaves of the tree, the most likely errors therefore correspond to scrambling only small patches of the input data.
We note that a good classifier should naturally be robust to small deformations
and corruptions of the input, and, in fact, adding various kinds of noise during
training is a commonly used strategy in classical machine learning.
Based on these intuitions, we expect our circuits could demonstrate a high
level of tolerance to noise.

In order to quantitatively understand the robustness of our proposed approach to
noise on quantum hardware, we study how performance is affected by
independent amplitude-damping and dephasing channels applied to each qubit. In
particular, we investigate how this error model would affect the pairwise tree network discriminative models of the type described in Section~\ref{sec:discriminative} and shown in Fig.~\ref{fig:d2_tree}.

The specific error model we implemented is the following: during the contraction step of node $i$ in the model evaluation, we compose amplitude damping and dephasing noise channels acting on its left and right children $\rho_{iL}$ and $\rho_{iR}$, mapping $\rho_{iL} \rightarrow \mathcal{E}_a\left(\mathcal{E}_d\left(\rho_{iL}\right)\right)$ and $\rho_{iR} \rightarrow \mathcal{E}_a\left(\mathcal{E}_d\left(\rho_{iR}\right)\right)$. 
Any completely positive trace-preserving noise channel $\mathcal{E}(\rho)$ can be expressed in the operator-sum representation as $\mathcal{E}(\rho) = \sum_a M_a \rho M_a^\dag$, where $\sum_a M_a M_a^\dag = I$. Here the Kraus operators $M_a$ for the amplitude damping channel $\mathcal{E}_a$ are (in the $z$-basis)
\begin{equation*}
M_0 = \begin{pmatrix} 1 & 0 \\ 0 & \sqrt{1-p_a} \end{pmatrix},\quad M_1 = \begin{pmatrix} 0 & \sqrt{p_a} \\ 0 & 0 \end{pmatrix},
\end{equation*}

while for the dephasing channel $\mathcal{E}_d$ the Kraus operators are

\begin{equation*}
M_0 = \sqrt{1-p_d}\, \mathbf{\mathit{I}}, \quad M_1 = \begin{pmatrix} \sqrt{p_d} & 0 \\ 0 & 0 \end{pmatrix},\quad M_2 = \begin{pmatrix} 0 & 0 \\ 0 & \sqrt{p_d} \end{pmatrix}.
\end{equation*}

To evaluate model performance under realistic values of $p_a$ and $p_d$ on current hardware, we determine $p_a$ and $p_d$ based on the continuous-time Kraus operators of these channels, which depend on the duration of the two-qubit gate $T_g$, the coherence time $T_1$ of the qubits, and the dephasing time $T_2$ of the qubits. Specifically, $p_a = 1 - e^{-T_g/T_1}$ and $p_d = 1 - e^{-Tg/T_2}$. Realistic
values for the time scales are $T_g = 200\,\mathrm{ns}$ and $T_1 = 50\,\mathrm{\mu s}$, $T_2 = 70\,\mathrm{\mu s}$, corresponding to $p_a = 0.004$ and $p_d = 0.003$. But numerical experiments with
these values showed almost no observable noise effects, so we consider an even more conservative parameter set with $T_1$ and $T_2$ reduced by an order of magnitude, such that $p_a = 0.039$ and $p_d = 0.028$. 

We plot the resulting test accuracies in Fig.~\ref{fig:triangle_plot_noisy}, noting that the Kraus operator formalism allows us to directly calculate the reduced density matrix of the labeling qubit under the effects of our noise model, therefore no explicit sampling of noise realizations is needed. 
Given that the coherence times used for the plot are easily achievable even on today's very early hardware platforms,
the results shown in Fig.~\ref{fig:triangle_plot_noisy} are encouraging: many of the models
give a test accuracy only slightly reduced from the noiseless case Fig.~\ref{fig:triangle_plot_noiseless}.
The largest reduction was for the digit `4' versus digit `9' model, which dropped from a test accuracy of  0.88 to 0.806. Interestingly this was also the model with the worst performance in the noiseless case.
The typical change in test accuracy across all of the models due to the noise was about 0.004.

To mitigate the effect of noise when classifying a particular image, one can evaluate the
quantum circuit some small number of times and choose the label which is
most frequently observed. For example, one could take a majority vote from 500
executions and classify and correctly classify an image whose individual
probability of success is \(.55\) with almost \(99\%\) accuracy.
In order to shed a more detailed light on our approach's robustness to noise, we plot in
Fig.~\ref{fig:noisy_evaluation} the individual success probabilities for classifying each
test example (\(x\)-axis), sorted by their probabilities for ease of visualization, 
over a range of decoherence times ($y$-axis).
The two panels show two different models, one trained to distinguish images of digits `0' versus `1' ;
the other digits `9' versus `4'.
These models were trained and evaluated at various levels of noise using the same training hyper-parameters that were found to give a good performance at $\{p_a,p_d\} = \{0.039,0.028\}$. The ratio between the dephasing time and the coherence time was held at a fixed ratio $T_2/T_1 = 7/5$.

We see that, for the examples which our models correctly classify in the low noise limit, the success probability remains appreciably greater than \(.5\) for a wide range of noise levels. 
In both diagrams the y-axis is scaled so that coherence times achievable by today's hardware occur two thirds of the way up from the bottom.
Interestingly, we note that the success probabilities saturate at coherence times much shorter than this, and only drop off dramatically at \(T_1\) values near $T_1 \sim 1 \,\mu\,s$.
The high performance of our model over a broad swath of tested coherence and dephasing times suggests that the effects of noise on our approach can be dramatically mitigated by the combination of the hybrid quantum/classical training procedure and a small number of repetitions with a majority voting scheme.
We find these results encouraging as empirical evidence that the limited-width ``causal cone" structure possessed by models of this type may have inherent noise robustness properties. 

\section{Discussion}

Many of the features that make tensor networks appealing for classical algorithms
also make them a promising framework for quantum computing. Tensor networks provide
a natural hierarchy of increasingly complex quantum states, allowing one
to choose the appropriate amount of resources for a given task. 
They also enable specialized algorithms which can make efficient use of valuable resources,
such as reducing the number of qubits needed to process high dimensional data.
An optimized, classically tractable tensor network  
can be used to initialize the parameters of a more powerful 
model implemented on quantum hardware. Doing so would alleviate issues associated with random
initial parameters, which can place circuits in regions of parameter space with vanishing gradients \cite{McClean:2018}.

While the approach to optimization we considered in our numerical experiments worked well,
algorithms which are more specialized to the tensor network architecture could be
devised. For example, by defining an objective for each subtree of a tree network
it could be possible to train subtrees separately \cite{Stoudenmire:2018}. 
Likewise, the MPS architecture has certain orthogonality or light-cone properties
which mean that only the tensors to the left of a certain physical index determine 
its distribution; this property could also be exploited for better optimization.

Another very interesting future direction would be to gain a better understanding
of the noise resilience of tensor network machine learning algorithms.
We performed some simple numerical experiments to show that these algorithms 
can tolerate a high level of noise, but additional empirical demonstrations as well 
as a theoretical explanation of how generic this property is would be very useful.
In an interesting recent work, Kim and Swingle investigated tensor networks
within a quantum computing framework for finding ground states of local 
Hamiltonians \cite{Kim:2017}.
One of their results was a rigorous bound on the sensitivity of the algorithm output to noise,
which relied on specific properties of tensor networks.
It would be very interesting adapt their bound to the machine learning context.

Other tensor network architectures besides trees and MPS also deserve further 
investigation in the context of quantum algorithms.
The PEPS family of tensor networks are specially designed to  
capture two-dimensional patterns of correlations \cite{Schwarz:2012,Schwarz:2013}.
The MERA family of tensor networks, retain certain benefits of tree tensor networks
but have more expressive power, and admit a natural description 
as a quantum circuit \cite{Vidal:2008,Kim:2017}.

Tensor networks strike a careful balance between
expressive power and computational efficiency, and can be viewed as a particularly useful
and natural class of quantum circuits. Based on the rich theoretical
understanding of their properties and powerful algorithms for optimizing them, 
we are optimistic they will provide many interesting avenues for quantum machine learning
research. \\

\section*{Acknowledgements}

We thank Dave Bacon, Norm Tubman, Alejandro Perdomo-Ortiz, Mark Nowakowski, Vinay Ramasesh, Dar Dahlen, and Lei Wang for helpful discussions. The work of W. Huggins and K. B. Whaley was supported by the U.S. Department of Energy, Office of Science, Office of Advanced Scientific Computing Research, Quantum Algorithm Teams Program, under contract number DE-AC02-05CH11231. Computational resources were provided by NIH grant S10OD023532, administered by the Molecular Graphics and Computation Facility at UC Berkeley.

\bibliography{main}

\end{document}